\documentclass[preprint2]{aastex6}
\usepackage{epsfig}
\usepackage{gensymb}
\usepackage{amsmath}
\usepackage{ulem}
\usepackage{verbatim}
\usepackage{commath}
\usepackage{mhchem}

\newcounter{chem}
\newcounter{temp}

\newenvironment{chequation}{%
  \setcounter{temp}{\value{equation}}%
  \setcounter{equation}{\value{chem}}%
}{%
  \setcounter{chem}{\value{equation}}%
  \setcounter{equation}{\value{temp}}%
}

\begin{document}

\title{Mechanism of Atomic Hydrogen Addition Reactions on np-ASW}

\author{Jiao He\altaffilmark{1} and Shahnewaj M. Emtiaz and Gianfranco Vidali\altaffilmark{2}}
\affil{Physics Department, Syracuse University, Syracuse, NY 13244, USA}
\altaffiltext{1}{jhe08@syr.edu}
\altaffiltext{2}{gvidali@syr.edu}

\shorttitle{mechanism of H addition reactions}
\shortauthors{He et al.}

\begin{abstract}
Hydrogen, being the most abundant element, is the driver of many if not most
reactions occurring on interstellar dust grains. In hydrogen atom addition
reactions, the rate is usually determined by the surface kinetics of the
hydrogen atom instead of the other reaction partner. Three mechanisms exist to
explain hydrogen addition reactions on surfaces: Langmuir-Hinshelwood,
Eley-Rideal, and hot-atom. In gas-grain models, which mechanism is assumed
greatly affects the simulation results. In this work, we quantify the
temperature dependence of the rates of atomic hydrogen addition reactions by
studying the reaction of H+O$_3$$\rightarrow$O$_2$+OH on the surface of a film
of non-porous amorphous solid water (np-ASW) in the temperature range from 10
K to 50 K. The reaction rate is found to be temperature independent. This
disagrees with the results of simulations with a network of rate equations that
assume Langmuir-Hinshelwood mechanism through either thermal diffusion or
tunneling diffusion; the reaction rates assuming such mechanism possesses
a strong temperature dependence, either explicitly or implicitly, that is not
seen experimentally. We suggest that the Eley-Rideal and/or hot-atom mechanism
play a key role in hydrogen atom addition reactions, and should be included in
gas-grain models. We also suggest that our newly developed time-resolved
reactive scattering can be utilized to measure the chemical desorption
efficiency in grain surface reactions.
\end{abstract}

\section{Introduction}
In the last decade, it has become increasingly evident that molecules that are
key to physical and chemical processes in the interstellar medium, such as
H$_2$, H$_2$O, H$_2$CO, CH$_3$OH and many others, are formed exclusively or in
large part on and in ices coating dust grains \citep{Herbst2009,Garrod2013}.
Evidence has come from observations \citep{Tielens2013}, laboratory experiments
\citep{Vidali2013}, and simulations \citep{Garrod2008}. In quiescent dense
clouds, where there is little UV flux, it can be argued that many of the
reactions are due to neutral atoms and radicals from the gas-phase impinging on
ice-coated dust grains and reacting with atoms and molecules on them
\citep{Linnartz2015}. How these reactions occur and at what rates is of
paramount importance in order to characterize and assess the role that dust
grains play in the chemical evolution of the ISM\@. One of the central
questions is the role of diffusion of the hydrogen atom that has landed on the
surface of a grain \citep{Biham2001, Cuppen2005, Cazaux2005, Hama2012,
Iqbal2012}. As an H atom approaches the surface of a dust grain, the H atom can
react with an atom or molecule on the surface through a direct hit (Eley-Rideal
mechanism), or it can make a few hops moving at super thermal speed and then
react (hot-atom mechanism). In both cases, there is little or no energy
accommodation with the surface. The third way of making a molecule is the
familiar Langmuir-Hinshelwood mechanism where the incoming hydrogen atom
becomes thermally accommodated with the surface. Hydrogen atoms move on the
surface either via thermal hopping, which obeys an Arrhenius type expression,
or via quantum tunneling, which is independent of surface temperature.
A gas-grain model assuming different mechanisms would yield dramatically
different results. Therefore, to characterize how reactions occur on surfaces
and obtain realistic prediction of the grain surface chemistry, one needs to
assess the mobility of hydrogen and reaction mechanisms. This type of
information is of great importance for simulations of the chemical evolution of
ISM environments, and is currently largely unavailable for systems of interest
to astrophysics. In this work, we focused our attention on measuring the
cross-section of the H+O$_3\rightarrow$O$_2$+OH reaction, where O$_3$ resides
on the surface of water ice, and deriving the mobility of atomic hydrogen on
the surface of water ice based on the cross-section \citep{Takahashi1999}.

In some reactions, the resulting molecule keeps a good fraction of the gained
bond energy and is ejected with high translational and ro-vibrational energy
\citep{Creighan2006,Gavilan2014}. Sometimes the ejection of the molecule due to
this type of non-thermal desorption caused by the reaction is called chemical
desorption \citep{Takahashi2000} to distinguish it from thermal desorption that
occurs when a particle leaves a surface because of its thermal energy. These
mechanisms have been verified and characterized in interaction of H(D) with
D(H) loaded single crystal surfaces, mostly metals \citep[e.g.][]{Jackson2002}.
Few experiments showing these two mechanisms have been carried out for
astrophysical relevant systems, mostly on graphitic \citep{Creighan2006,
Gavilan2014} and polycyclic aromatic hydrocarbons (PAHs) surfaces
\citep{Mennella2012}. Chemical desorption of H$_2$O due to the reaction of
oxygen and hydrogen on silicate was detected by \citet{Dulieu2013}.
\citet{Minissale2016} reported an extensive list of chemical desorption
efficiencies of various grain surface reactions; however, no detailed
supporting experimental data were provided. In this study we use a newly
developed method --- time-resolved reactive scattering --- to measure
{\underline{directly}} the chemical desorption efficiency of
H+O$_3$$\rightarrow$O$_2$+OH and demonstrate how to utilize this technique to
measure the chemical desorption efficiency. We then carry out simulations to
show that the Eley-Rideal/hot-atom mechanism has to be incorporated in
simulations of atom addition reactions on grain surfaces.

The paper is structured as follows. The apparatus is described in the next
Section. In Section~\ref{sec:exp} the experimental methods, results and their
analysis are presented. The significance of these measurements for
characterizing chemical reactions on surfaces is given in
Section~\ref{sec:astro}.

\section{Apparatus}
\label{sec:apparatus}
Experiments were performed in a ultra-high vacuum (UHV) chamber
\citep{He2015b,He2016b}. At the center of the 10 inch diameter stainless steel
chamber there is a 1 cm$^2$ gold coated copper disk that can be cooled down to
8~K using liquid helium or heated up to 450~K using a cartridge heater. The
temperature of the sample disk is measured and controlled by a Lakeshore 336
temperature controller with an accuracy better than 0.05~K. A Hiden Analytical
quadrupole mass spectrometer (QMS) mounted on a rotatable platform can directly
face one of the molecular beams to measure beam composition, or face the sample
to measure molecules coming off the sample. A gas capillary array is placed
behind the sample holder to deposit water vapor into the chamber for ice
growth. In this study, experiments were performed either on a clean gold surface
or non-porous amorphous solid water (np-ASW). The growth of water ice has
been described in \citet{He2016b}; here we briefly summarize it. The np-ASW
sample was grown by background deposition of water vapor onto the gold
substrate when the substrate was at 130~K. The water vapor pressure during
deposition was $5\times 10^{-7}$ torr, and the deposition duration was 200 s,
amounting to about 100 Langmuir of water ice. After water deposition, the ice
sample was annealed at 130~K for 30 minutes, during which the pressure in the
chamber drops to middle $10^{-10}$ torr range. Water ice prepared this way
is np-ASW \citep{Stevenson1999}. Connected to the main reaction chamber are two
three-stage differentially pumped atomic/molecular beamlines. In this study only
one beam was used. The first stage of the beamline houses a radio-frequency (RF)
powered dissociation source with an inductive coil wrapped around a water
cooled Pyrex glass tube. The end of the source was capped by an aluminum nozzle
with an inner diameter of 1~mm. Under the feeding pressure we used, the beam
was effusive with a Boltzmann velocity distribution characterized by
a temperature of $\sim$300~K. An Alicat MCS-5 mass flow controller was used to
accurately control the gas flow to the dissociation source. Gas specific
correction factors are already taken into account by the flow controller. The
beam flux was calibrated using temperature programmed desorption (TPD)
experiments \citep{He2016b}. For experiments performed in this study, the H$_2$
and O$_2$ flow were $1.4\times 10^{-2}$ ML/s and $5.6\times10^{-3}$ ML/s,
respectively. The uncertainty in flow rate is mostly due to the uncertainty in
determining the monolayer coverage in the TPD experiments, and it is estimated
to be about 30\%. When the RF power was turned on, the dissociation rates for
H$_2$ and O$_2$ were about 70\% and 30\%, respectively. Based on a previous
measurement of the speed of the atomic beam, atoms from the source are well
thermalized with the Pyrex glass wall \citep{He2016a}. The second stage of the
beamline houses an in-vacuo DC motor driving a chopper disk with a single open
slit with a 1/40 duty factor. For experiments performed in this study, the
chopper spinning speed was set to 50$\pm1$ Hz. A pair of LED and photodiode
located at the opposite sides of the chopper disk was used to monitor the
opening and closing of the beam. The pulse signal generated by the photodiode
when the 1/40 open slit was lined up with the LED-photodiode pair was fed into
a multichannel scaler (MCS) coupled with the QMS\@. This provide capability to
measure the in-phase intensity of the beam or the molecules desorbing from the
surface. During a typical in-phase measurement in this work, the dwell time of
the MCS, which is also the time resolution of the time-of-flight spectra, was
set to 20 $\mu$s, and a total of 10,000 or 20,000 scans were averaged to
increase the signal-to-noise ratio. Considering the period of the chopper disk,
20 ms, and the duty cycle, 1/40, the equivalent H exposure time, 5 to 10
seconds, results in a small dose to the sample. The consumption of ozone by
H was only a small fraction of a monolayer. For simplicity, in later
discussions we assume that the ozone coverage was always 1 ML during the
in-phase measurements. More details of the in-phase detection can be found in
\citet{He2016a}. In the third stage of the beamline, a stepper motor controlled
flag automates the opening and closing of the beam. A LabVIEW program controls
the stepper motor so that the uncertainty in the beam open time is less than 50
ms.

\section{Experiments, Results and Analysis}
\label{sec:exp}
\subsection{Preparation of the ozone layer}
\begin{figure}[ht!]
  \epsscale{1}
  \plotone{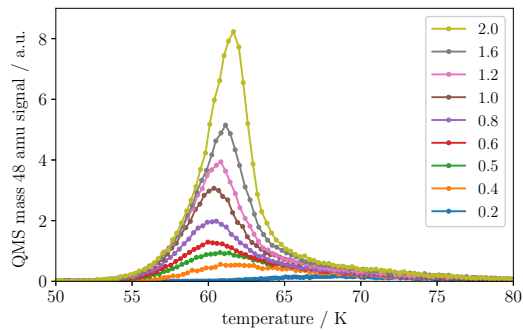}
  \caption{Temperature programmed desorption (TPD) traces of ozone on a gold
  surface. Heating ramp rate during TPD was 0.5 K/s. Surface coverage is shown
in the inset in the unit of monolayer (ML).}
\label{fig:ozone_TPD}
\end{figure}

The ozone layer was prepared following a procedure as described in
\citet{He2014c}. With the substrate at 20~K, the dissociated oxygen beam was
sent onto the sample. The atomic oxygen and undissociated molecular oxygen
react and form ozone. The sample was then brought up to 50~K and annealed at
50~K for a few minutes to remove molecular oxygen, leaving ozone on the
surface. In certain conditions, a small fraction (less than $\sim$1\%) of
atomic oxygen may still be present \citep{He2015b}. We will ignore the effect
of atomic oxygen because it is small compared to other uncertainties. The ozone
was then cooled down to a lower temperature for further experiments. A complete
set of ozone TPD experiments with different O/O$_2$ deposition durations were
performed on a gold surface. The TPD traces are shown in
Figure~\ref{fig:ozone_TPD}. At low coverages, ozone molecules tend to occupy
deep adsorption sites and desorb at relatively high temperatures. As the
coverage increases from 0.2 ML to 1.0 ML, deeper sites are filled gradually and
ozone molecules occupy shallower sites with lower desorption temperatures. The
decrease in desorption peak temperature is accompanied by a decrease in the gap
between the leading edges of successive traces decreases. Above 1 ML, the
desorption is zeroth order with the typical overlapping leading edges. We
determine the one monolayer coverage based on the gap of the leading edges.
This method of thickness calibration is the same  as the one used in several
other works~\citep{He2014b,Smith2016,He2016b}. The calibration of TPD
experiments of different molecules such as O$_2$ and CH$_4$ (not shown here)
demonstrated that the number of adsorption sites per unit area of np-ASW is
similar to that of gold coated copper disk used as sample. This suggests that
it takes the same amount of O/O$_2$ exposure to cover np-ASW with one ML of
O$_3$ as to cover the gold surface. Therefore the calibration of ozone coverage
on the gold surface applies to the np-ASW surface.

\subsection{Time-resolved reactive scattering in H+O$_3$ experiments}
In the H+O$_3 \rightarrow \rm{O}_2$+OH reaction, depending on the surface
temperature, a fraction of the reaction product O$_2$ leaves the surface via
either thermal desorption or chemical desorption. The thermal desorption has
the well-known rate given by the Arrhenius-type expression $\nu \exp(-E_{\rm
des}/k_{\rm B} T)$, where $\nu$ is the desorption pre-factor, $E_{\rm des}$ is
the desorption energy (or binding energy), $k_{\rm B}$ is the Boltzmann
constant, and $T$ is the surface temperature. Thermal desorption has a strong
dependence on the temperature. Chemical desorption occurs when the heat
released in the reaction makes the product (O$_2$ in this case) leave the
surface. Usually the chemical desorption rate is insensitive to the surface
temperature because the energy from the reaction is much higher than the
binding energy, especially for weakly bounded species like O$_2$. Therefore, by
studying the reaction at different surface temperature we should be able to
distinguish the two mechanisms of desorption. At relative high surface
temperature ($T>35$~K) the thermal desorption rate of O$_2$ is high, and all
O$_2$ leave the surface within a short time, whether thermally or chemically.
In H+O$_3$ experiments carried out at low surface temperatures ($T<25$~K), the
thermal desorption rate is negligible but the chemical desorption rate should
remain the same as that of high temperature. The chemical desorption
efficiency, which is defined as the fraction of reaction products that leave
the surface due to the exothermicity of the reaction, can be approximately
calculated as the desorption yield at low temperature divided by the desorption
yield at high temperature.

To determine the desorption rate of O$_2$ accurately, we used an in-phase
detection, the time-resolved reactive scattering (TRRS) method, because the
traditional method of using the QMS is inadequate for the measurement of the
chemical desorption rate. Typically a QMS has a time scale (dwell time) of
a few hundred milliseconds, so it is not possible to separate chemically
desorbed molecules from those thermally desorbed with a residence time
comparable with the QMS time scale. To overcome this difficulty of conventional
QMS measurement, we used a modulated atomic hydrogen beam and multichannel
scaler (MCS) coupled with the QMS to measure the in-phase signal of O$_2$, so
that the O$_2$ that leaves the surface due to chemical desorption can be
separated from the thermally desorbed O$_2$. The in-phase detection also
increase the sensitivity of the measurements because the O$_2$ contribution from
the chamber background is subtracted out. In addition, the sample holder, which
is also at low temperature, may act as a cryopump with a changing pumping
speed. This affects the O$_2$ measured by the QMS\@. In-phase detection
overcomes these problems. A more detailed description of the advantage of
in-phase detection can be found in \citet{He2016a}.

\begin{figure}[ht!]
  \epsscale{1}
  \plotone{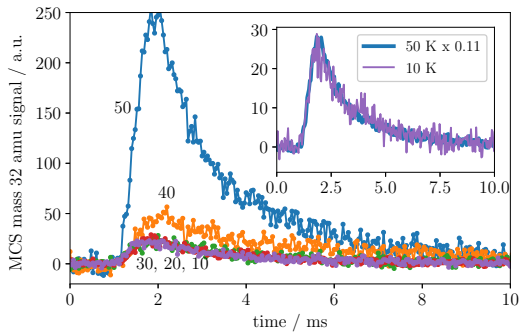}
  \caption{Multichannel scaler signal of mass 32 amu during the exposure of
  pulses of atomic hydrogen beam on 1 ML of ozone on np-ASW. The signal is
  averaged over 10000 or 20000 cycles. The surface temperature is shown in the
  figure. The inset shows a comparison between the trace at 50 K and that at 10
  K.}
\label{fig:mcs_H_O3}
\end{figure}

To obtain the temperature dependence of reaction cross-section of H+O$_3
\rightarrow \rm{O}_2$+OH, atomic hydrogen was deposited on top of a full layer
of ozone at 10, 20, 30, 40, and 50~K. In the in-phase measurements, to ensure
that H+O$_3$ experiments done at different temperatures start from the same
initial condition, the initial surface for each experiment was fully covered
with one monolayer of ozone, and the subsequent H exposure was kept short.
Based on prior discussion, the change in ozone coverage during H exposure is
a small fraction of one monolayer. Therefore an incoming H atom always sees
a whole layer of ozone, and the underlying substrate does not play a role in
the reaction. The resulting time-resolved reactive scattering of O$_2$ from
H+O$_3$ is shown in Figure~\ref{fig:mcs_H_O3}. The x-axis shows the time delay
respect to the time when the chopper slit is aligned up with the beam, and the
y-axis shows the averaged mass 32 amu pulse counts within 20~$\mu$m bins. 
The spread of the peak in Figure~\ref{fig:mcs_H_O3} is mostly due to the time
of flight from the chopper to the sample. The distance from the chopper to the
sample is much longer than the distance from the sample to the QMS detector. In
our previous paper \citep{He2016a}, we reported that the spread of the peak is
consistent with an effusive beam at room temperature with Maxwell-Boltzmann
distribution. The measured mass 32 amu signal can also be due to other
reaction products, such as HO$_2$ and H$_2$O$_2$. For now, we assume that the
mass 32 amu signal is solely due to O$_2$. Later, we'll show that other
contributions to mass 32 amu signal only amount to a small fraction of the
O$_2$ signal. At 50~K, the O$_2$ desorption is the highest. From 50 to
30~K, the desorption of O$_2$ drops dramatically because of the change in the
thermal desorption rate. At 30, 20, and 10~K, the mass 32 amu desorption rates
are almost the same, indicating that the thermal desorption rate is zero and
the desorption is exclusively chemical. The inset shows the mass 32 amu
desorption rate at 10~K is 0.11 times as the desorption rate at 50~K. Therefore
the chemical desorption rate is 11\% of the total desorption rate of O$_2$. We
only carried out the H+O$_3$ measurement at up to 50~K, because at higher
surface temperatures ozone starts to desorb during the experiment. We do not
exclude the possibility that at 50~K, there is thermal desorption of a small
fraction of O$_2$ with a residence time longer than the time scale of MCS
measurement (a few ms). These molecules would show up as background and be
subtracted out, therefore an underestimation of the total O$_2$ desorption is
possible. We conclude that the chemical desorption rate of O$_2$ produced by
H+O$_3$ could be slightly less than 11\% of the total desorption rate. This
value is close to the measurement by \citet{Minissale2016} which found the
upper limit to be 10\% and 8\% on amorphous silicate and oxidized HOPG,
respectively.

\subsection{Measurement of the H+O$_3$ cross section}
\begin{figure}[ht!]
   \epsscale{1}
   \plotone{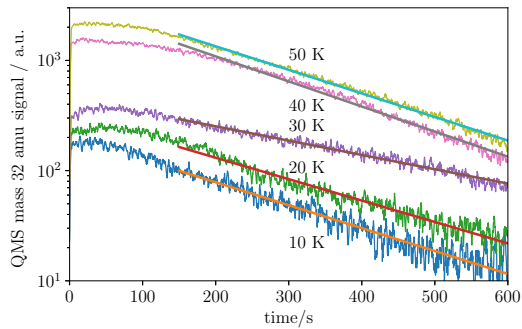}
   \caption{The mass 32 amu signal measured by the QMS during atomic hydrogen
	   exposure on 1 ML of ozone at 10, 20, 30, 40, and 50 K. The hydrogen
	   atom beam flux is about $2\times10^{-2}$~ML/s. The straight lines
	   are fits using an exponential decay law, see text. }
\label{fig:exp_decay}
 \end{figure}

In order to measure the reaction cross-section of H+O$_3$, we carried out
another set of experiments without in-phase detection. We irradiated the ozone
sample at 50, 40, 30, 20, and 10~K with H for an extended period of time until
the ozone was used up. In each of these measurements, 1 ML of ozone was
prepared on the np-ASW surface before atomic hydrogen deposition. The ozone
sample was irradiated until the desorption of O$_2$ became negligible. The QMS
was used to monitor the decay of O$_2$, see Figure~\ref{fig:exp_decay}. At the
very beginning of H exposure, the surface is fully covered with O$_3$ and the
reaction rate does not drop with time. It also suggests that the initial ozone
coverage might be slightly more than 1 ML\@. When the ozone coverage becomes
less than 1 ML the reaction rate decays exponentially. The traces for different
temperatures, plotted in a semi-log scale, are fitted well with straight lines.
We confirmed that the measured mass 32 amu signal is due to the reaction
H+O$_3$ instead of air contamination in the H/H$_2$ beam by aligning the QMS
detector entrance with the beam and measuring the beam composition directly.  

The following reactions are likely to occur on the surface:

\begin{chequation}
\begin{align}
  \ce{H} + \ce{O3} &\rightarrow \ce{O2} + \ce{OH} \label{r:h_o3}\\
  \ce{H} + \ce{OH} &\rightarrow \ce{H2O} \label{r:h_oh} \\
  \ce{H} + \ce{O2} &\rightarrow \ce{HO2} \label{r:h_o2}\\
  \ce{H} + \ce{HO2} &\rightarrow \ce{H2O2} \label{r:h_ho2}\\
  \ce{H} + \ce{H2O2} &\rightarrow \ce{H2O} + \ce{OH} \label{r:h_h2o2}\\
  \ce{H} + \ce{H} &\rightarrow \ce{H2}   \label{r:h_h}
\end{align}
\end{chequation}

At 50~K and 40~K, almost all of the O$_2$ that is formed leaves the surface
within a very short time. O$_2$ does not build up on the surface.
Reactions~\ref{r:h_o3} and~\ref{r:h_oh} dominate. At 10~K and 20~K, most of
O$_2$ remains on the surface and can react with atomic hydrogen
via~\ref{r:h_o2}.  Almost all of the above reactions are involved. In
Figure~\ref{fig:exp_decay}, notice that the desorption at 40~K is lower than
that of 50~K. This difference is likely due to the fact that the effective
pumping speed of the sample holder is higher when the sample is at 40~K than at
50~K \citep{He2016a}. At lower temperatures the traces are much lower, both
because of the difference in pumping speed and a lower thermal desorption rate
of mass 32 amu at lower temperatures. The exponential decay in O$_2$ is similar
at all temperatures except at 30~K. The anomalous behavior at 30~K can be
explained by the coverage dependence of O$_2$ binding energy. At 50 and 40~K,
no O$_2$ builds up on the surface, and the thermal desorption rate is not
changing with time. At 30~K, at the beginning of the H exposure, the O$_2$
coverage is low and the binding energy is high, and the thermal desorption rate
is low. As O$_2$ builds up, O$_2$ has to occupy sites with progressively weaker
binding energy and therefore the thermal desorption of O$_2$ increases. More
details on the coverage dependence of binding energy can be found in
\citet{He2016b}. Therefore, there is significant change in slope of the O$_2$
decay at 30~K with respect to at 40 and 50 K. At 20~K and 10~K, O$_2$ does not
thermally desorb and stays on the surface except for those O$_2$ that
chemically desorb. O$_2$ reacts with atomic hydrogen, leading to the products,
HO$_2$ and H$_2$O$_2$; they could all be possible sources of mass 32 amu in the
QMS as shown in Figure~\ref{fig:exp_decay}, because they can be ionized to
O$_2^+$ in the QMS. To calculate the reaction cross-section of H+O$_3$, the
contribution from the other two molecules has to be subtracted. This is
discussed next.

\begin{figure}[ht!]
  \epsscale{1}
  \plotone{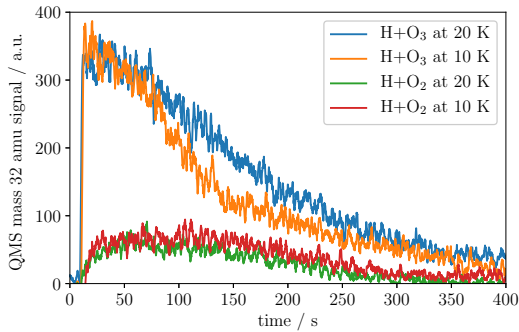}
  \caption{Mass 32 amu signal during H exposure on 1.7 ML of O$_2$ (bottom two
  traces) and 1 ML of ozone (top two traces) at 10 K and 20 K. The measurements
were performed on an np-ASW sample. }
\label{fig:H_O3_O2}
\end{figure}

To find out how reactions~\ref{r:h_o2} and~\ref{r:h_ho2} contribute to the mass
32 amu signal in the QMS, we studied H+O$_2$ and H+O$_3$ reactions under the
same conditions. An np-ASW surface was covered with 1.7 ML of O$_2$ or 1 ML of
O$_3$, and then exposed to the H beam until the O$_2$/O$_3$ was consumed and
mass 32 amu signal dropped to the background level. The measurements were
carried out both at 10~K and 20~K. The mass 32 amu signal was recorded during
the H exposure, as shown in Figure~\ref{fig:H_O3_O2}.  We excluded the
possibility that H$_2$ (which is present as the undissociated fraction in the
beam) sputters off O$_2$ ice by sending an H$_2$ beam onto O$_2$ ice. No O$_2$
sputtered off the sample was detected. The only possibility left to explain the
mass 32 signal during H exposure to O$_2$ is that HO$_2$ desorbs and is
detected in the QMS as mass 32 amu (O$_2^+$).  HO$_2$ most likely (partly)
chemically desorbs from the surface and is efficiently broken up in the QMS
ionizer. From the comparison between H+O$_2$ and H+O$_3$, one can see that in
the same condition, H+O$_2$ only introduces $\sim$20\% of mass 32 amu as that
of H+O$_3$. If we take the chemical desorption efficiency of
reaction~\ref{r:h_o3} as 0.11, then the chemical desorption efficiency for
reaction~\ref{r:h_o2} is $0.2\times0.11=0.022$. It should be noted that we are
assuming the detection efficiency of HO$_2$ (as mass 32 amu) and O$_2$ are the
same in the QMS\@. But even without this assumption, a similar conclusion for
the H+O$_3$ reaction cross-section can be reached, because the chemical
desorption rate of \ref{r:h_o2} is trivial. It should be also noted that in
H+O$_3$ the mass 32 amu signal increases as soon as the H exposure begins,
while in H+O$_2$ the mass 32 amu signal does not jump to maximum right away.
This delay may be attributed to a small reaction energy barrier in
reaction~\ref{r:h_o2}. Quantifying the barrier of reaction~\ref{r:h_o2} is out
of scope of this work and we assume it is barrierless hereafter. In both
H+O$_2$ and H+O$_3$ experiments, we also checked other masses, and no
significant changes in mass 17, 18, 33, and 34 amu signals were seen during
H exposure. Based on our previous measurements, if H$_2$O$_2$ desorbs, mass 34
amu should be seen from the QMS signal.  Non-detection of mass 34 during
H exposure suggests that H$_2$O$_2$ does not contribute significantly to the
mass 32 amu signal. The mass 32 amu signal is mostly due to the desorption of
O$_2$, and a small contribution from HO$_2$ most likely because it partly
chemically desorbs from the surface and is efficiently broken up in the QMS
ionizer.  

\subsection{Rate equation simulations}
We used a simple set of rate equations to describe the interaction of H with
O$_3$. For atomic hydrogen addition reactions, we consider all
three possibilities: Langmuir-Hinshelwood, quantum tunneling, and
Eley-Rideal/hot-atom mechanisms. To avoid complications, we only consider one
mechanism at a time to see the effect of each of them. Based on Reactions
\ref{r:h_o3} to \ref{r:h_h}, we come up with the following rate equations to
describe the reaction network:

\begin{widetext}
\begin{eqnarray}
     \Psi=   \begin{cases}
     \phi \sigma_{\rm ER},  &\text{E-R or H-A} \\
   \nu N(\mathrm{H}) \exp (-\frac{E_{\rm diff, H}}{k_{\rm B}T}),  &\text{L-H}\\
   \nu N(\mathrm{H}) \exp (-2a \sqrt{2m_{\rm H}E_{\rm diff, H}}/\hbar),  &\text{tunneling}
    \end{cases}\\
  \frac{\dif N(\mathrm{O}_3)}{\dif t} =-\Psi N(\mathrm{O}_3)\\
  \frac{\dif N(\mathrm{O}_2)}{\dif t} =(1-R_1)\Psi N(\mathrm{O}_3)- \nu N(\mathrm{O}_2)\exp (-\frac{E_{\rm des, O_2}}{k_{\rm B}T}) -\Psi N(\mathrm{O}_2)  \\
  \frac{\dif N(\mathrm{HO}_2)}{\dif t} =(1-R_2)\Psi N(\mathrm{O}_2) -\Psi N(\mathrm{HO}_2)  \\
  \frac{\dif N(\mathrm{H_2O}_2)}{\dif t} =\Psi N(\mathrm{HO}_2)- \Psi \exp(-\frac{E_{\rm a}}{k_{\rm B}T}) N(\mathrm{H_2O_2})\\
  \frac{\dif N(\mathrm{H_2O})}{\dif t} =\Psi N(\mathrm{OH})+ \Psi \exp(-\frac{E_{\rm a}}{k_{\rm B}T}) N(\mathrm{H_2O_2})  \\
  \frac{\dif N(\mathrm{OH})}{\dif t} =\Psi N(\mathrm{O}_3)+ \Psi \exp(-\frac{E_{\rm H_2O_2}}{k_{\rm B}T}) N(\mathrm{H_2O_2}) -\Psi N(\mathrm{OH})  \\
  \frac{\dif N(\mathrm{H})}{\dif t} =\phi -\nu N(\mathrm{H})\exp (-\frac{E_{\rm des, H}}{k_{\rm B}T})-\Psi N(\mathrm{H_2O_2}) \exp({-\frac{E_{\rm a}}{k_{\rm B}T}})- \nonumber \\ \Psi(2N(\mathrm{H})+N(\mathrm{O}_3)+N(\mathrm{O}_2)+N(\mathrm{HO}_2)+N(\mathrm{OH}))
\end{eqnarray}
\end{widetext}

$N_{xx}$ are the coverage of different species on the surface, and they have
the dimensionless-unit of ML. $\Psi$ is the term linked to the rate of
atomic hydrogen addition reactions.  $\phi$ is the incoming atomic hydrogen
flux.  $\sigma_{\rm ER}$ is the reaction cross-section in the E-R or H-A
mechanism.  Since both the E-R  and the H-A mechanisms follow similar numerical
expressions, except that they have different values of the reaction
cross-section, we treat them indifferently in the rate equation simulations.
The chemical desorption efficiency of reactions~\ref{r:h_o3} and~\ref{r:h_o2}
are represented by $R_1$ and $R_2$, respectively. The chemical desorption of OH
from H+O$_3$ was checked by measurement, and no significant OH desorption was
seen using the QMS\@. In Reaction~\ref{r:h_h2o2}, the reaction energy barrier
$E_{\rm a}$ is taken to be 2100~K \citep{Koussa2006}. All other reactions are
taken to be barrierless \citep{Cuppen2010, Mokrane2009}. The thermal desorption
of O$_2$ is calculated by assuming a single value, 1200K, of the binding
energy, although ideally a continuous distribution should be used instead
\citep{He2016b}. The binding energy of H on AWS is taken to be 450
K \citep{Garrod2006}. Different values of $\alpha=E_{\rm diff}/E_{\rm des}$
ratio were tried in the simulation. But only simulation results with $\alpha=
0.7$ are presented as an illustration. The desorption of HO$_2$, H$_2$O, and
H$_2$O$_2$ are ignored in the simulation because there is negligible desorption
of these molecules during the H exposure.

Before presenting the simulation results, we do a qualitative analysis of the
reaction cross-section. Based on Figure~\ref{fig:exp_decay}, the exponential
decay rate of mass 32 amu signal is similar at 10, 20, 40, and 50 K. Little
temperature dependence is seen. Therefore the value of $\Psi$ should have
little temperature dependence between 10 K and 50 K. In the expression for L-H
mechanism, a strong temperature dependence is expected because of the Arrhenius
type expression. In the expression of diffusion by tunneling there is no
explicit temperature dependence; however, the coverage of hydrogen atoms on the
surface $N(\mathrm{H})$ strongly depends on the surface temperature. Therefore
the only mechanisms that do not possess temperature dependence are the E-R and
H-A mechanisms.  \citet{Yuan2014} explained temperature
independent reaction rate of OH+CO$\rightarrow$CO$_2$+H by E-R mechanism in
a similar way.

In the simulations assuming the L-H mechanism, different values of the
H binding energy and diffusion energy barrier were tried, but none  were
able to yield exponential decay curves with little temperature dependence, as
observed experimentally. The results of one of the simulations are shown in
Figure~\ref{fig:sim_LH}. The L-H mechanism is unable to explain the measured
data in Figure~\ref{fig:exp_decay}. Similarly, we also confirmed that quantum
tunneling diffusion mechanism is unable to produce temperature independent
exponential decay curves as in Figure~\ref{fig:exp_decay}. The
Eley-Rideal/hot-atom mechanism has to be used instead. The cross-section in the
E-R/H-A mechanism that best reproduces the traces in Figure~\ref{fig:exp_decay}
is shown in Table~\ref{tab:cross-section}. The cross-section is calculated in
the unit of the number of square lattices. The third column is calculated using
an adsorption sites density of $1.5\times 10^{15}$ cm$^{-2}$, although
different densities have been assumed in previous models. The relative errors
of the values of the cross-section are mostly due to the uncertainty in fitting
and the fluctuation of the atomic hydrogen flux, and they are estimated to be
less than 5\%. However, the absolute error, which is mostly due to the
uncertainty in determining the surface coverage, can be as large as 30\%.  The
cross-sections at different temperatures are close with little temperature
dependence; and they are smaller than the area of a square lattice, indicating
that H does not move much on the surface.

\begin{figure}[ht!]
  \epsscale{1}
  \plotone{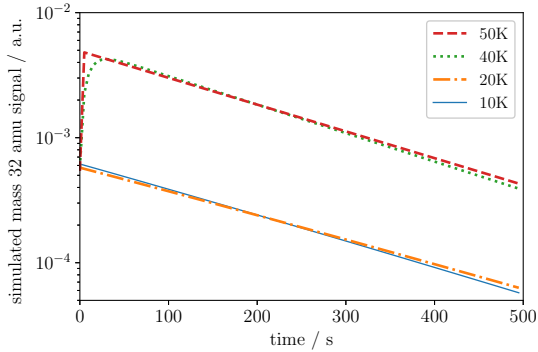}
  \caption{Simulated mass 32 amu signal using the rate equations model, assuming
    the Eley-Rideal/hot-atom mechanism. The parameters used in the simulation
    are shown in Table~\ref{tab:para}. }
\label{fig:sim_ER}
\end{figure}

\begin{figure}[ht!]
  \epsscale{1}
  \plotone{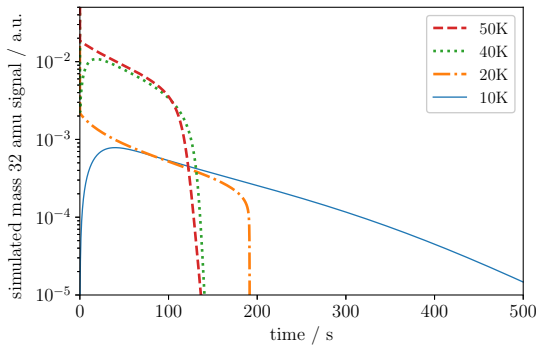}
  \caption{As in Figure~\ref{fig:sim_ER}, but assuming the Langmuir-Hinshewood
  mechanism.}
\label{fig:sim_LH}
\end{figure}

\begin{table}
\caption{Cross-section of the reaction H+O$_3\rightarrow$O$_2$+OH on np-ASW at
  different temperatures calculated using a rate equation model assuming
  Eley-Rideal/hot-atom mechanism for atomic hydrogen. The second column is
  the cross-section in the unit of $l_0^2$, where $l_0$ is the distance between
  two nearest adsorption sites. The cross-section in the third column is
  calculated assuming an adsorption site density of $1.5\times 10^{15}$
  cm$^{-2}$. }
\label{tab:cross-section}
\centering
\begin{tabular*}{0.4\textwidth}{@{\extracolsep{\fill} }ccc}
temperature & cross-section/l$_0^2$ & cross-section/\AA$^2$ \\
       & \\ \cline{1-3}
    &\\
10        & 0.28       & 1.89                         \\
20        & 0.27       & 1.77                         \\
40        & 0.27       & 1.77                         \\
50        & 0.25       & 1.66

\end{tabular*}
\end{table}

\begin{table}
\caption{Parameters in the rate equation simulation for results in Figure~\ref{fig:sim_ER} and Figure~\ref{fig:sim_LH}.}
\label{tab:para}
\centering
\begin{tabular*}{0.3\textwidth}{@{\extracolsep{\fill} }cc}
Parameter & Value  \\
     \cline{1-2}    &\\
$E_{\mathrm{des,H}}$      & 450~K                     \\
$E_{\mathrm{des,O}_2}$        & 1200~K         \\
$E_{\mathrm{diff,H}}$      & 315~K                     \\
$E_{\mathrm{a}}$      & 2100~K                     \\
$\nu$        & $10^{12}$ s$^{-1}$                   \\
$\phi$        & 0.020 ML$\cdot$s$^{-1}$      \\
$R_1$        & 0.11      \\
$R_2$        & 0.022
\end{tabular*}
\end{table}
\section{Astrophysical Implications}
\label{sec:astro}
Reactions on surfaces of dust grain analogs involving hydrogen atoms are the
dominant reactions that govern grain surface chemistry. How hydrogen atoms take
part in the reactions, and specifically what is the mechanism for the diffusion
of hydrogen atoms, are important issues in astrophysics. Here, in this work we
demonstrated that the reaction between hydrogen atoms and ozone --- a venue to
form water on surfaces \citep{Tielens1982} --- cannot be explained by
using the Langmuir-Hinshelwood mechanism, either with thermal diffusion or quantum
tunneling diffusion. Eley-Rideal or hot-atom type mechanisms have to be involved to
interpret the experimental data. Although in our experiment we started with
a full layer of ozone on np-ASW, the conclusion that the reaction is E-R is
based on the analysis of the H+O$_3$ reaction over the entire coverage range of
ozone. Our conclusions regarding the mechanisms of reactions involving hydrogen
atoms can be generalized to other barrierless and low-barrier hydrogen atom
addition reactions on grain surfaces, because the reaction rate is governed by
the kinetics of hydrogen atoms instead by the one of ozone or other reactants.
We studied the reaction with H atoms at thermal energy ($\sim$300K), while
H atoms in certain environments --- such as in dense clouds --- may have lower
kinetic energies, and it is possible that the cross-section values have to be
modified. But the mechanism of the reaction should still be the same. In the case
that the reaction has a barrier, the rate needs to be multiplied by $exp(-E/k_{\rm
B}T)$, where $E$ is the reaction energy barrier. In most existing gas-grain
models, the Langmuir-Hinshelwood mechanism is the main mechanism considered, if
not the only mechanism, although there are now models incorporating Eley-Rideal
or hot-atom mechanisms to study the formation of molecules in the ISM
\citep{Ruaud2015}. For hydrogen atoms, quantum tunneling diffusion needs also
to be considered, although it depends greatly on the morphology of the surface.
The main result of our work is that our laboratory measurements and simulations
demonstrate the importance of Eley-Rideal and/or hot-atom mechanisms. Together
with the other mechanisms, they have to be incorporated in models of surface
reactions with hydrogen atoms.

\section{Acknowledgments}
We would like to thank Eric Herbst and Ilsa Cooke for their helpful suggestions
and Xixin Liang for technical assistance. We are grateful to an
anonymous referee for the constructive comments that helped to improve the
paper. This work was supported by NSF Astronomy \& Astrophysics Research
Grants No.1311958 and No.1615897.
\bibliography{H_O3}
\bibliographystyle{apj}
\end{document}